\documentclass[onecolumn,prd,amsmath,amsssymb,groupedaddress,12pt]{revtex4}
\usepackage{graphicx}
\usepackage{dcolumn}
\usepackage{bm}
\usepackage{amssymb}
\def\lsim{\mathrel{\mathstrut\smash{\ooalign{\raise2.5pt\hbox{$<$}\cr\lower2.5pt\hbox{$\sim$}}}}}
\def\gsim{\mathrel{\mathstrut\smash{\ooalign{\raise2.5pt\hbox{$>$}\cr\lower2.5pt\hbox{$\sim$}}}}}
\def\be{\begin{equation}}
\def\ee{\end{equation}}
\def\bea{\begin{eqnarray}}
\def\eea{\end{eqnarray}}

\begin{document}

\title{Unveiling Chameleons in Tests of Gravitational Inverse-Square Law}
\author{Amol Upadhye$^{1}$, Steven S. Gubser$^{1}$, and Justin Khoury$^{2}$}
\date{July 28, 2006}

\affiliation{$^1$ Department of Physics, Princeton University, Princeton, NJ 08544, USA \\
$^2$ Perimeter Institute for Theoretical Physics, 
31 Caroline St. N., Waterloo, ON, N2L 2Y5, Canada}

\begin{abstract}
Scalar self interactions are known to weaken considerably the current constraints on scalar-mediated fifth forces.  We consider a scalar field with a quartic self interaction and gravitation-strength Yukawa couplings to matter particles.  After discussing the phenomenology of this scalar field, we assess the ability of ongoing and planned experiments to detect the fifth force mediated by such a field.  Assuming that the quartic and matter couplings are of order unity, the current-generation E\"ot-Wash experiment at the University of Washington will be able to explore an interesting subset of parameter space. The next-generation E\"ot-Wash experiment is expected to be able to detect, or to rule out, the fifth force due to such a scalar with unit quartic and matter couplings at the 3$\sigma$ confidence level.
\end{abstract}

\maketitle

\section{Introduction} 

Years of effort have been devoted to searching for new macroscopic forces from sub-millimeter to solar system scales~\cite{fischbach}. From a theoretical standpoint, modern theories of particle physics introduce new scalar fields which can mediate long range forces. This is certainly true of string theory, whose plethora of moduli generically couple to matter with a strength comparable to that of gravity. Provided they remain light, these scalars should cause observable deviations from the gravitational inverse square law and violations of the equivalence principle.

\begin{figure} 
\includegraphics[width=5in]{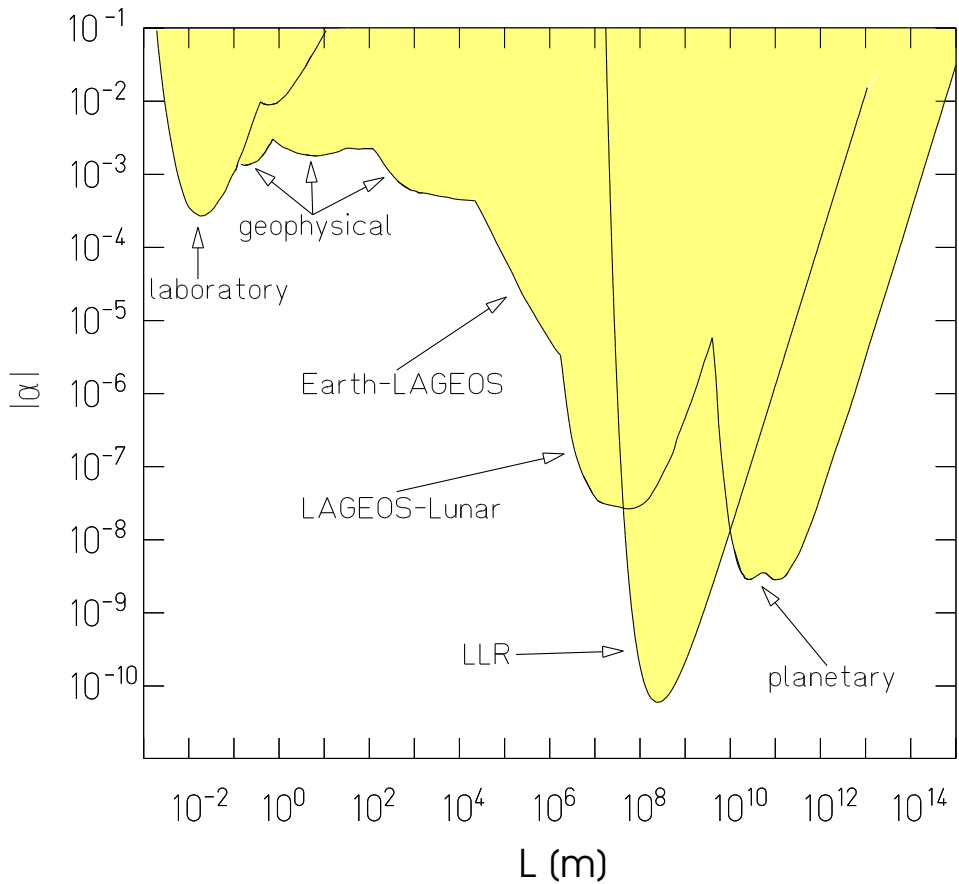}
\caption{Current experimental constraints on the strength $\alpha$ and range $L$ (in meters) of a Yukawa fifth force, ignoring self-interactions. Reprinted from~\cite{adellong}.}
\label{expcons} 
\end{figure}

The experimental state of affairs is shown in Fig.~\ref{expcons}.  Evidently, a fifth force of gravitational strength, $\alpha\sim {\cal O}(1)$, is excluded on all scales ranging from 0.1~mm to $10^2$~AU. A crucial underlying assumption, however, is that the mediating scalar field has negligible self-interactions. As argued in~\cite{gubserkhoury}, the addition of a quartic term drastically changes the picture. Gravitational-strength coupling to matter is allowed again, even with a quartic coupling as small as $\sim 10^{-53}$~\cite{nelson}. The ability for self-interacting scalars to hide from experiments relies on the chameleon mechanism~\cite{original,gubserkhoury}, which suppresses fifth-force signals in two ways.

Firstly, the presence of ambient matter density generates a tadpole term in the Klein-Gordon equation, which shifts the minimum of the potential. Because of the quartic coupling, the mass of small fluctuations around this effective minimum can be much larger than the mass in vacuum. This is most emphatically illustrated with a massless field with quartic coupling of order unity: in a medium of density 1~g/${\rm cm}^3$, the effective mass is $0.1$~mm. 

Secondly, the fifth force is further suppressed by the thin shell effect, another startling consequence of the nonlinearity of the field equations. Within a dense macroscopic body, the effective mass of the scalar is large. As a result, the contribution of the core to the external field is exponentially small.  Only a thin shell near the surface exerts a significant pull on an exterior test particle. In fact, for an infinite plate, the force at the surface eventually saturates as the thickness is increased, causing the strength relative to gravity to fall off rapidly.

It follows that an ideal experiment to detect a chameleon-like scalar must: 1.~use sufficiently small test masses to minimize the thin shell suppression; and 2.~probe the force at distances $\lsim 1$~mm to avoid the exponential damping from the effective mass. These considerations point towards the E\"ot-Wash experiment at the University of Washington~\cite{hoyle}.

In this paper we carefully assess the ability of the E\"ot-Wash experiment to detect or exclude chameleon scalar fields with quartic self-interactions. 
We find that the current apparatus with two disks of 42 holes each is not sensitive enough to detect a chameleon force with dimensionless quartic coupling of order unity, $\lambda\sim {\cal O}(1)$, and gravitational strength coupling to matter, $\beta =1$. It should nevertheless place significant constraints on models with larger values of $\beta$ or smaller $\lambda$. The next generation E\"ot-Wash experiment, whose design involves two disks with 120 wedges removed, is much more promising for detecting self-interacting scalars. Our calculations show it should detect or exclude chameleon fields with $\lambda=\beta=1$ at the 3$\sigma$ level. Unfortunately it looks unlikely that the next generation E\"ot-Wash will be able to distinguish between a chameleon-mediated force and a Yukawa force with suitable strength and range.

We begin in Sec.~\ref{phenom} with a summary of the phenomenology of self-interacting scalar fields. While it is well-known that free scalars mediate attractive forces, in Sec.~\ref{attract} we argue that this is also the case for chameleon scalars. The chameleon mechanism and thin-shell effect are reviewed in Secs.~\ref{subsec:chameleonmech} and~\ref{subsec:thinshell}, respectively. We describe in Sec.~\ref{predicts} the expected fifth force signals for self-interacting scalars for the current and next generation E\"ot-Wash experiments. Section~\ref{conclu} summarizes our results and discusses prospects for distinguishing chameleon-mediated forces from Yukawa forces.


\newcommand\TL{\hfil$\displaystyle{##}$}
\newcommand\TR{$\displaystyle{{}##}$\hfil}
\newcommand\TC{\hfil$\displaystyle{##}$\hfil}
\newcommand\TT{\hbox{##}}
\newcommand\JOT{\noalign{\vskip1\jot}}
\def\seqalign#1#2{\vcenter{\openup1\jot
  \halign{\strut #1\cr #2 \cr}}}
\def\lbldef#1#2{\expandafter\gdef\csname #1\endcsname {#2}}
\newcommand{\eqn}[3][]{\lbldef{#2}{(\ref{#2})}%
\def\@eqnstyle{#1}%
\ifx\@eqnstyle\@empty%
\begin{equation} \eqalign{#3} \label{#2} \end{equation}%
\else%
\begin{equation} \seqalign{\span\TC}{#3} \label{#2} \end{equation}%
\fi}
\def\eqalign#1{\vcenter{\openup1\jot
    \halign{\strut\span\TL & \span\TR\cr #1 \cr
   }}}
\def\eno#1{(\ref{#1})}

\section{Phenomenology of the self-interacting scalar} \label{phenom}

\subsection{Attractive force theorem} \label{attract}

An intuitively appealing conjecture is that scalar-mediated forces between identical objects should always be attractive.  Certainly this is true when the scalar is a free field.  But is it true when the scalar can interact with itself and/or other fields?  A simple analytical argument gives some support for this conjecture.  To start, consider an action functional for a single scalar field of the form
 \eqn{OneScalar}{
  S[\phi] = \int d^4 x \, \left[ \frac{1}{2} G(\phi)
    \partial_\mu \phi \partial^\mu \phi - V(\phi) \right] -
    \sum_\alpha \int_{\gamma_\alpha} m_\alpha(\phi) ds \,,
 }
where the sum is over all particles (e.g.~atomic nuclei) which couple to the scalar.  The integrals inside this sum are over the world lines $\gamma_\alpha$ of these particles.  The coupling of the scalar to each particle is through some $\phi$ dependence of its mass.  This dependence is typically very weak (i.e.~Planck suppressed), so it is quite a mild assumption to assert that $m_\alpha(\phi) > 0$ everywhere.  The functions $G(\phi)$ and $V(\phi)$ are also assumed to be positive everywhere, except that $V(\phi)$ must be $0$ at its unique global minimum.  Positivity of $G(\phi)$ is guaranteed by unitarity, and positivity of $V(\phi)$ is required for vacuum stability.  Now consider two parallel plates that are uniform and infinite in the $y$ and $z$ directions and of arbitrary but finite thickness in the $x$ direction.  The plates are assumed to be identical, so that they may be exchanged by reflecting through the plane midway between them.  Initially, let us say the plates are touching.  Then we move each of them a distance $a/2$ away from the other, so that the reflection symmetry is always through the plane $x=0$.  We wish to show that the total energy is an increasing function of the separation $a$: this is what it means for the force between the plates to be attractive.

The total energy per unit area for a given separation distance $a$ and a given static scalar field configuration $\phi = \phi(x)$ is
 \eqn{TotalE}{
  \frac{H[\phi,a]}{A} = \int_{-\infty}^\infty dx \, \left[ \frac{1}{2} G(\phi)
    (\partial_x \phi)^2 + V(\phi) + U_1(\phi,x-a/2) + 
    U_2(\phi,x+a/2) \right] \,.
 }
Here $U_1$ and $U_2$ account for the interactions of the plates with the scalar: if $A$ is the area of the plates, then when $a=0$ we have
 \eqn{GotpOne}{
  U_i(\phi,x) A = \sum_{\alpha \in i} m_\alpha(\phi)
   \delta(x-x_\alpha) \,,
 }
where $x_\alpha$ is the $x$-coordinate of the position of the $\alpha$-th particle.  The sum in \eno{GotpOne} runs over all the particles in plate $i$.  The claim is that
 \def\min{\mathop{\rm minimum}}
 \eqn{IsIncreasing}{
  H(a)/A \equiv \min_\phi H[\phi,a]/A
 }
is an increasing function of $a$.  To see this, consider $a_1 < a_2$.  Assume that the minimum of $H[\phi,a_2]$ is attained for $\phi = \phi_2(x)$.  Now consider the following test function:
 \eqn{TestFct}{
  \phi_1^*(x) = \left\{ \eqalign{
    \phi_2\left( x + {a_2-a_1 \over 2} \right) &\quad
     \hbox{for $x > 0$}  \cr
    \phi_2\left( x - {a_2-a_1 \over 2} \right) &\quad
     \hbox{for $x < 0$} \,.
  } \right.
 }
\begin{figure}[tb]
\includegraphics[width=3in]{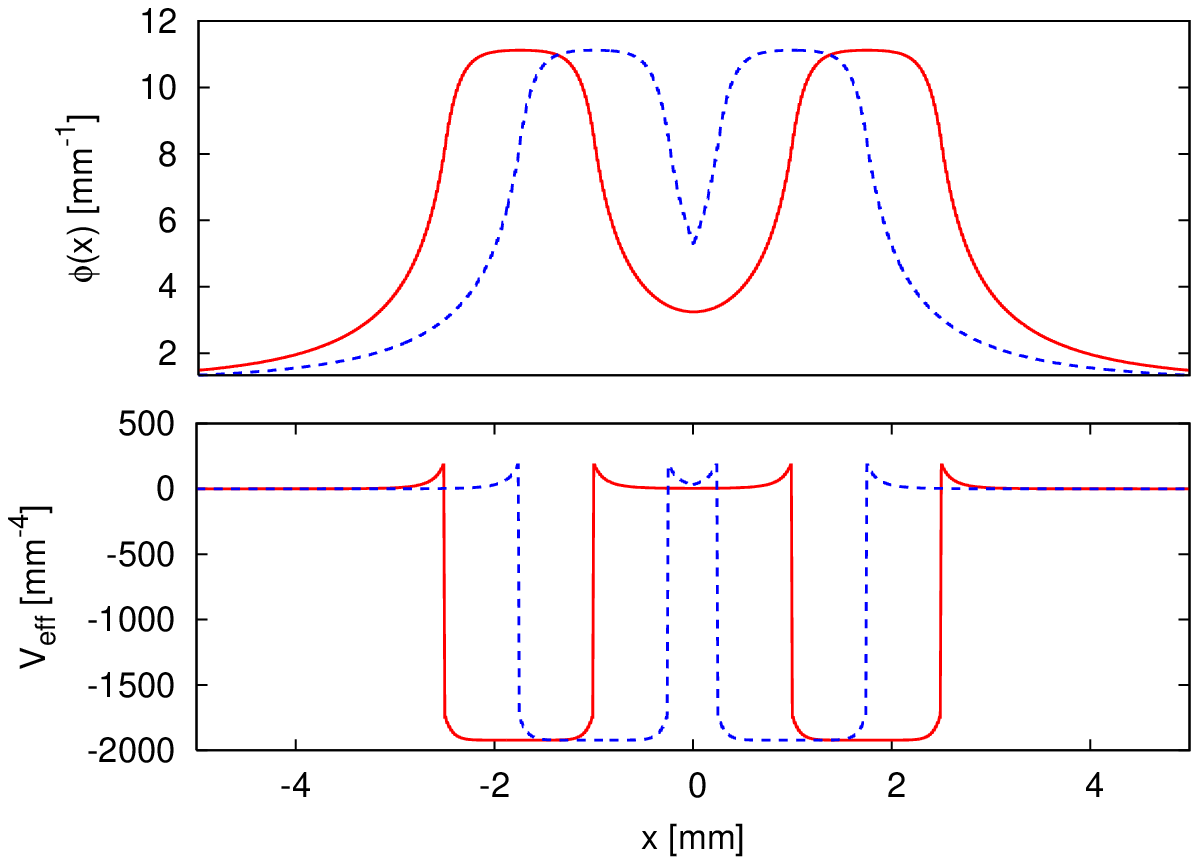}
\includegraphics[width=3in]{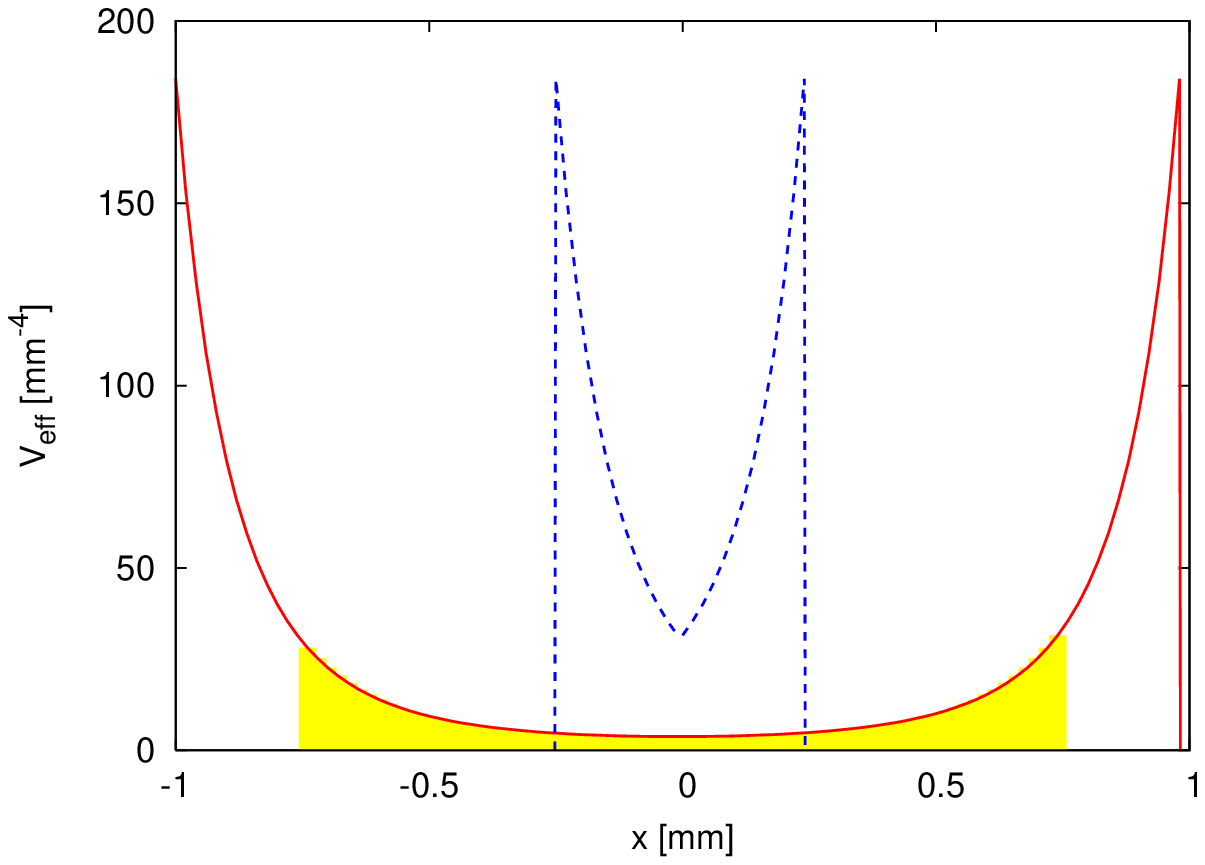}
\caption{(upper left) The solid line shows the field $\phi_2(x)$ for two plates, for a particular choice of the potential, and the dashed line shows the spliced-together field $\phi_1^*(x)$.  (lower left) The effective potential $V_{\rm eff}=V(\phi)+U_1(\phi,x)+U_2(\phi,x)$ is shown for each of these fields.  (right) A closeup of the $V_{\rm eff}$ plot shows that the effective potential is positive in the region that is cut out.  The area under the curve in this region, shaded in the plot, is the difference in energies between the two field configurations $\phi_2$ and $\phi_1^*$.\label{fig:attractV}}
\end{figure}In words, we form $\phi_1^*$ by cutting out the center region $\left( -{a_2-a_1 \over 2},{a_2-a_1 \over 2} \right)$ of the minimizer $\phi_2$ for separation $a_2$.  This procedure is illustrated in Fig.~\ref{fig:attractV}(left).  Because of the assumption of reflection symmetry through the $x=0$ plane, $\phi_1^*$ is a continuous function, but its first derivative flips sign at $x=0$.  Now we reason that
 \eqn{Ecutout}{
  \frac{H(a_1)}{A} \leq \frac{H[\phi_1^*,a_1]}{A} = 
   \int_{-\infty}^{-(a_2-a_1)/2} dx \, {\cal H}_2(x) + 
   \int_{(a_2-a_1)/2}^\infty dx \, {\cal H}_2(x) \leq 
   \frac{H[\phi_2,a_2]}{A} = \frac{H(a_2)}{A} \,,
 }
where
 \eqn{CalEDef}{
  {\cal H}_2(x) \equiv {1 \over 2} G(\phi_2)
    (\partial_x \phi_2)^2 + V(\phi_2) + U_1(\phi_2,x-a_2/2) + 
    U_2(\phi_2,x+a_2/2) \,
 }
is the Hamiltonian density for the field configuration $\phi_2(x)$.
The first inequality in \eno{Ecutout} follows simply from noting that $\phi_1^*$ is probably not the minimizer $\phi_1$ for separation $a_1$.  The next equality comes from using \eno{TestFct}.  The next inequality follows from having ${\cal H}_2(x) \geq 0$ everywhere.  The final equality follows from the construction of $\phi_2$.
In other words, the region between the two plates that we cut out has a positive energy, as shown in Fig.~\ref{fig:attractV}(right).  Removing this region lowers the total energy.

This ``cutting out the middle'' argument can be generalized to include more complicated source masses, as well as several scalars.  Gauge fields can also be included with couplings to the scalars, provided the matter in the source masses does not couple to them in any way.  The validity of the argument we have given depends on the following assumptions:
 \begin{enumerate}
  \item Quantum effects, such as zero-point energy, can be neglected.
  \item The energy density is bounded from below for all possible configurations, and is positive in the region between the masses.
  \item There exists a unique vacuum field configuration.\label{item:Vacuum}
  \item The two source masses are mirror images under the reflection symmetry through the $x=0$ plane.\label{item:Reflection} 
  \item Terms with derivatives higher than first order can be neglected.\label{item:FirstDerivatives}
  \item The minimizing functions, e.g.~$\phi_2$ in \eno{TestFct}, must exist: that is, the minimum of $H[\phi,a]/A$ must be attained for any fixed $a$.  
\label{item:Attain}
 \end{enumerate}
Point~\ref{item:Vacuum} precludes the possibility of domain wall configurations of the scalar field(s), which are undesirable for various phenomenological reasons, and which could also spoil the argument we have given by interfering with the reflection symmetry.

It should be possible to replace the assumption in Point~\ref{item:Reflection} with a milder one, namely that both masses should have the same sign of scalar charges (meaning that the derivatives of their masses with respect to the scalar have the same signs).  More ambitiously, one might hope to show that two or more objects without reflection symmetry exert only attractive forces on each other via couplings to scalar fields.  But in such generalizations, it clearly becomes much less trivial to choose a good test function analogous to $\phi_1^*$.

Point~\ref{item:FirstDerivatives} is important because the test function $\phi_1^*(x)$ usually has a jump in its first derivative at $x=0$.  If $H[\phi,a]/A$ involved $\phi''$, then there could be a delta function contribution to $H[\phi_1^*,a]/A$ which is not captured by the second equality in \eno{Ecutout}.

The importance of Point~\ref{item:Attain} can be appreciated by considering the case of a free massless scalar with $G(\phi)=1$ and $m(\phi) = m + e\phi$.  Then the energy functional \eno{TotalE} is precisely what we would obtain from electrostatics, where $\phi$ is the voltage and $e$ is the electric charge.  This presents an apparent paradox: we know that two positively charged plates repel, but naive application of the argument \eno{Ecutout} indicates that the plates attract.  The resolution of course is that Point~\ref{item:Attain} is important.  A correct treatment of the electrostatic case includes planar ``screening charges'' at $x=\pm\infty$, the sum of whose charges exactly cancels the sum of the charges of the plates; otherwise the energy functional is not gauge-invariant.  (Or, if one wishes to avoid gauge theoretic concepts, one could argue that without the screening charges, the scalar is drawn into runaway behavior by the like-signed plates.)  If the two screening charges are equal, respecting the reflection symmetry, then the voltage is constant between the plates and has linear slope elsewhere, corresponding to constant electric fields pointing outward from the plates to infinity.  The energy functional is infinite, but if it is regulated by bringing the screening charges in a little from infinity to an $a$-independent position, then it is simple to see from the energy functional that the plates indeed repel.  Another way to say the same thing is that they are drawn outward by their attraction to the screening charges \footnote{One should not conclude from this discussion that a massless scalar field would cause a repulsion between finite identical plates; indeed, such a field would cause an attractive force.  The apparent tension with the discussion in the main text is an order of limits issue: the first limit we took in the argument \eno{TotalE}-\eno{CalEDef} is that the plates were infinite in extent, and this limit interacts oddly with the zero-mass limit for the scalar.}.

\subsection{Chameleon mechanism}\label{subsec:chameleonmech}
From now on, we choose a specific scalar field theory, with a mass term and a $\phi^4$ self interaction,
\begin{equation}
V(\phi) = \frac{1}{2} m_\phi^2 \phi^2
	+ \frac{\lambda}{4!}\phi^4\,.
\label{eqn:V_chameleon}
\end{equation}
As in \cite{gubserkhoury}, the masses of matter particles are assumed to be only weakly dependent on $\phi$, 
\begin{equation}
m_\alpha(\phi) = m_\alpha\left(1 + \frac{\beta\phi}{M_{\rm Pl}}\right)\,,
\label{eqn:m_alpha}
\end{equation}
where the constant $\beta$ is assumed to be the same for all matter particles.  The matter action in \eno{OneScalar} can then be written
\begin{equation}
S_{\rm matter}[\phi] = -\sum_\alpha \int_{\gamma_\alpha} m_\alpha(\phi) ds
 	= - \int d^4x \rho(\vec x)\left(1 
			+ \frac{\beta\phi(\vec x)}{M_{\rm Pl}}\right)\,,
\label{eqn:S_matter}
\end{equation}
leading to the effective potential
\begin{equation}
V_{\text{eff}}(\phi,\vec x) = \frac{1}{2} m_\phi^2 \phi^2
	+ \frac{\lambda}{4!} \phi^4 - \frac{\beta}{M_{\rm Pl}} \rho(\vec x) \phi\,,
\label{eqn:Veff_chameleon}
\end{equation}
once terms independent of $\phi$ have been discarded.  Note that the last term on the right hand side of (\ref{eqn:Veff_chameleon}) becomes a Yukawa coupling term $-\frac{\beta m}{M_{Pl}} {\bar \psi}\psi \phi$ in the case of nonrelativistic Fermionic matter.  The equation of motion resulting from (\ref{eqn:Veff_chameleon}) is
\begin{equation}
-\partial_\mu \partial^\mu \phi = m_\phi^2 \phi + \frac{\lambda}{3!}\phi^3 - \frac{\beta}{M_{\rm Pl}} \rho(\vec x) = \frac{dV_{\text{eff}}}{d\phi}\,.
\label{eqn:eom}
\end{equation}

In a uniform material of nonzero density $\rho$, the minimum of the effective potential is shifted to positive values of $\phi$, and the scalar field picks up an effective mass
\begin{equation}
m_{\text{eff,}\rho}^2 = \left. \frac{d^2 V_{\text{eff}}(\phi)}{d\phi^2} \right|_{\phi_\rho}
	 = m_\phi^2 + \frac{1}{2} \lambda \phi_\rho^2\,,
\label{eqn:meff}
\end{equation}
where $\phi_\rho$ is the field value which minimizes the effective potential.
This density dependence of the effective mass, known as the chameleon mechanism, allows the field to ``hide'' by decreasing its interaction length in the presence of matter.  The effect is most pronounced for a massless scalar, $m_\phi=0$, which would mediate a long-range interaction in the absence of the self interaction term.  In this case, the effective length scale is given by
\begin{equation}
m_{\text{eff,}\rho}^{-1} = \left(\frac{2}{9}\right)^{1/6} \beta^{-1/3} \lambda^{-1/6} \left(\frac{M_{\rm Pl}}{\rho}\right)^{1/3}\,.
\label{eqn:Leff}
\end{equation}
For $\lambda=\beta=1$ and $\rho = 1$ g/cm$^3$, this length scale is $0.13$ mm.  That is, the fifth force becomes negligible compared to gravity for an object of this density at distances much larger than one tenth of a millimeter.

As an example of the chameleon mechanism, consider the field near an infinite plate of uniform density and nonzero thickness, surrounded by a vacuum.  This case is interesting because, for a long range interaction such as gravity, the force due to such a plate is independent of distance from the plate.  Furthermore, the force due to a Yukawa interaction is known to fall off exponentially with distance from the plate (see, {\it e.g.}, \cite{adellong}).

The equation of motion of the chameleon field simplifies to
\begin{equation}
\frac{d^2\phi}{dx^2} = \frac{\lambda}{3!} \phi^3\,,
\label{eqn:eom_plate}
\end{equation}
where $m_\phi=0$, and the $x$ direction is assumed to be normal to the plate.
The appropriate vacuum solution, satisfying the boundary condition $\phi(\infty)=0$ at positive infinity, is
\begin{equation}
\phi(x) = \frac{\sqrt{12/\lambda}}{x-b}\,.
\label{eqn:phi_vac}
\end{equation}
Here, the parameter $b$ is determined by the matching condition at the surface of the plate.  The solution diverges at $x=b$, so this point must be either inside or on the other side of the plate.  We note that for $x\gg b$, $\phi(x)$ is proportional to $\lambda^{-1/2}$ and is independent of $\beta$ as well as the density of the source plate, consistent with the results of \cite{nelson}.

The chameleon-mediated fifth force felt by a test particle sitting outside the plate is 
\begin{equation}
F_c(x) = \int  d^3x \; \beta \rho_{\text{test}} \nabla \phi = \beta m_{\text{test}} \nabla\phi
= \frac{-\beta m_{\text{test}} \sqrt{12/\lambda}}{(x-b)^2}\,.
\label{eqn:F_test}
\end{equation}
It is evident that, although the gravitational force on the test particle remains constant with increasing $x$, the fifth force falls off as $x^{-2}$ far from the plate.  The self interaction transforms the fifth force due to $\phi$ from a long-range interaction into a short-range one.

\begin{figure}[tb]
\includegraphics[width=3in]{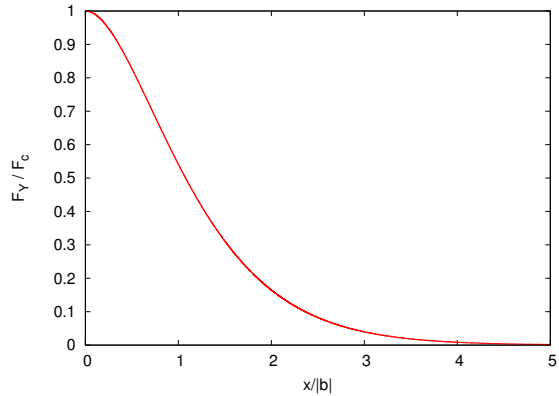}
\caption{Ratio of the Yukawa force to the chameleon force due to an infinite plate of matter.  The strength and length scale of the Yukawa force have been chosen so as to match the chameleon force and its first derivative at the surface of the plate, $x=0$. \label{fig:y_vs_c}}
\end{figure}

Another useful comparison is between the chameleon force and the Yukawa force, as shown in Fig.~\ref{fig:y_vs_c}.  The strength and characteristic length of the Yukawa force $F_Y(x)$ have been chosen such that $F_Y(0)=F_c(0)$ and $F_Y'(0) = F_c'(0)$, where the surface of the plate is assumed to be at $x=0$.  Figure~\ref{fig:y_vs_c} shows that $F_Y(x)$ falls off much more rapidly than $F_c(x)$ at distances $x \gtrsim |b|$.  It will be shown in Sec.~\ref{subsec:thinshell} that, for a sufficiently thick plate, $F_c(x)$ is independent of the plate thickness.  In this case, the only remaining length scale in the problem is $m_{\text{eff,}\rho}^{-1}$, so by dimensional analysis, $|b| \sim m_{\text{eff,}\rho}^{-1}$.  Therefore, the Yukawa force becomes substantially weaker than the chameleon force at distances of the order of the chameleon mass scale inside the test mass.

\subsection{Thin shell effect}\label{subsec:thinshell}

For a massless scalar, $m_\phi=0$, $V_{\text{eff}}$ has a minimum $dV_{\text{eff}}/d\phi=0$ only if the field has a self interaction, $\lambda \neq 0$.  That is, for $m_\phi=\lambda=0$, in an infinitely large material, $\phi$ will rise without bound, just like the gravitational potential in an infinitely large, uniform-density sphere.  The self-interaction cuts off this increase at some maximum field value $\phi_\rho$, within a few scale lengths $m_{\text{eff,}\rho}^{-1}$ of the edge of the material.  As a result, only the thin shell of material near this edge contributes to the field value outside the material.  Since the fifth force felt by a test particle outside the material is proportional to the gradient of the field, the test particle will only feel a fifth force from the thin shell of material, rather than the bulk; this is known as the thin shell effect.

\begin{figure}[tb]
\includegraphics[width=5in]{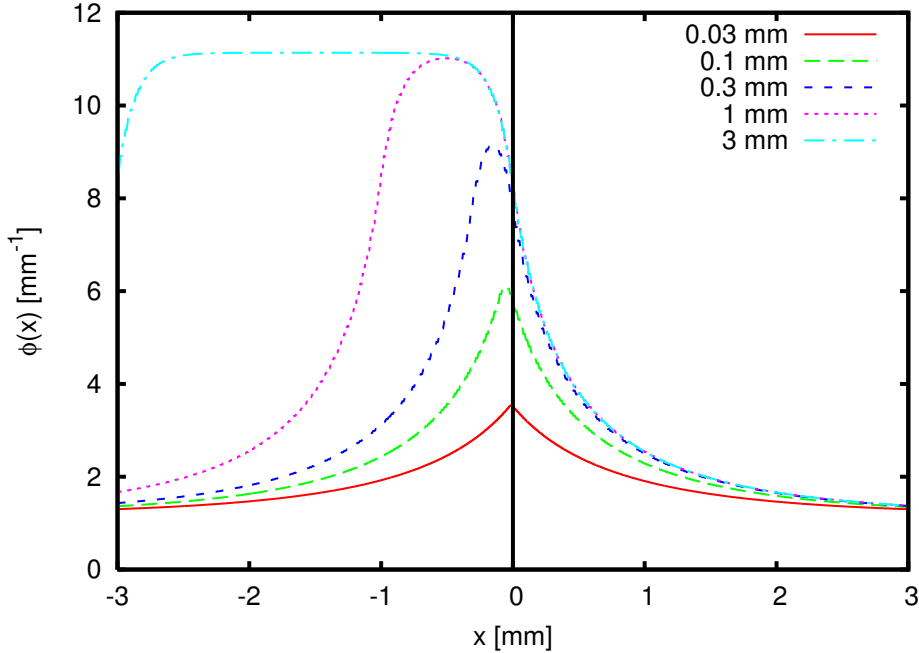}
\caption{$\phi(x)$ for several infinite plates, of various thicknesses, lined up so that their rightmost edges are at $x=0$.  The density inside each plate is $1$ g/cm$^3$, and the density outside is $10^{-3}$ g/cm$^3$.  \label{fig:thinshell}}
\end{figure}

\begin{figure}[tb]
\includegraphics[width=5in]{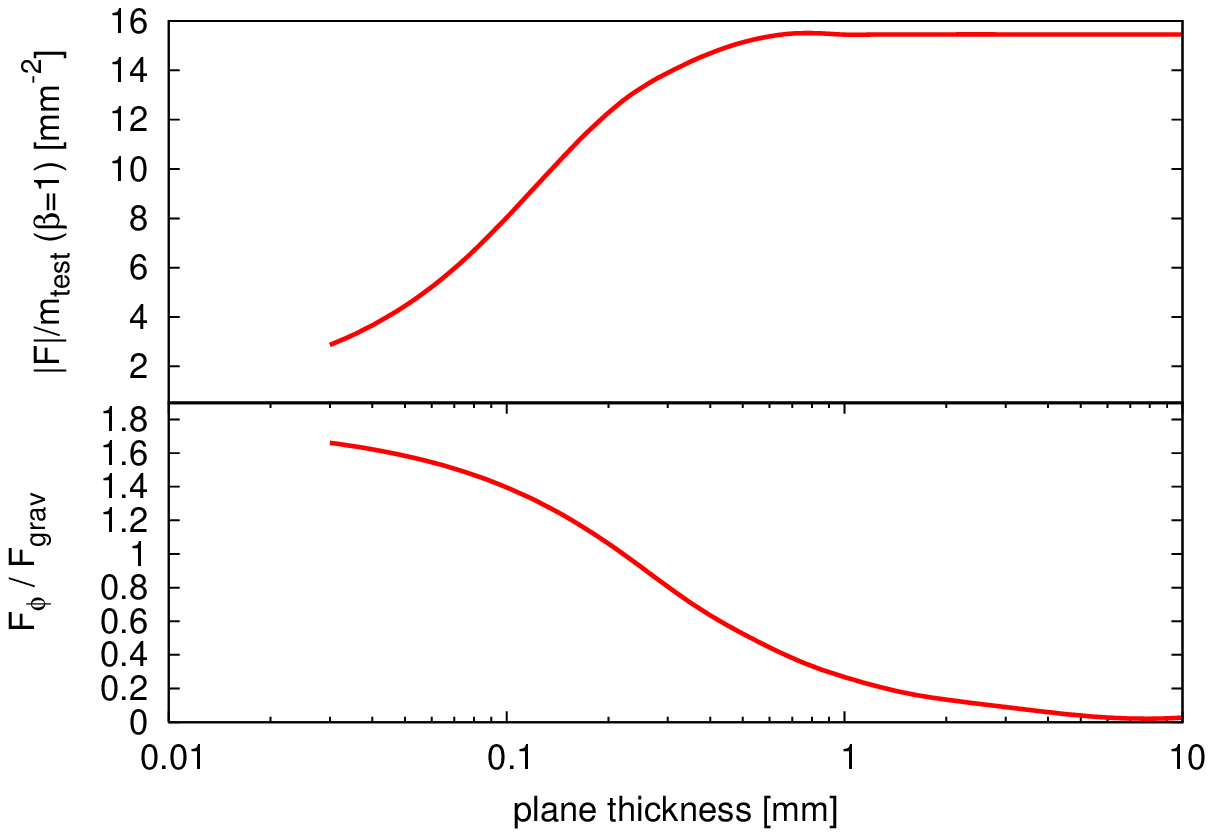}
\caption{Saturation of the fifth force.  (top) The force on the test particle per unit mass approaches a maximum value as the plate thickness exceeds $m_{\text{eff,}\rho}^{-1}$.  (bottom) The gravitational force on the test particle continues to grow, so that the fifth force is only a small correction to gravity for large objects.  \label{fig:saturate}}
\end{figure}

The thin shell effect is illustrated in Fig.~\ref{fig:thinshell}.  Evidently, once the thickness of the plate has grown to a few times the scale length $m_{\text{eff,}\rho}^{-1}$, any further increases in thickness leave the field outside the plate essentially unaffected.  Since the fifth force on a test particle is proportional to the gradient of the field, the fifth force saturates for plate thicknesses a few times $m_{\text{eff,}\rho}^{-1}$.  Meanwhile, the gravitational force on the test particle continues to grow linearly with the plate thickness, causing the ratio of the fifth force $F_\phi$ to the gravitational force $F_{\text{grav}}$ to fall off rapidly, as shown in Fig.~\ref{fig:saturate}.

\subsection{Applicability of the uniform density approximation}
The above discussions of the phenomenology of the self interacting scalar field have treated the sources of the field as objects of uniform density.  Since actual matter consists of a lattice of atoms or molecules, and since the equation of motion (\ref{eqn:eom}) of the self-interacting scalar is nonlinear, one may question the validity of approximating matter as a substance of uniform density.  Ref. \cite{mota} shows that this approximation is indeed valid, provided that the length scale $m_{\text{eff,}\rho}^{-1}$ is much larger than the interatomic separation.  The weak dependence of $m_{\text{eff,}\rho}^{-1}$ on $\beta$ means that this condition is only violated for $\beta$ many orders of magnitude greater than unity.

\subsection{Summary of chameleon properties}
We have seen that a self-interacting scalar field with a gravitation-strength Yukawa coupling to matter tends to give rise to an attractive force.  In a medium of density $\rho$, it acquires a length scale $m_{\text{eff,}\rho}^{-1}$, turning the chameleon-mediated fifth force into a short range interaction.  Furthermore, the chameleon force between two objects much larger than $m_{\text{eff,}\rho}^{-1}$ couples only to a thin outer shell on each object.

Such a field is particulary difficult to observe.  The chameleon fifth force falls off rapidly with distance, so an experiment must be able to test gravity at separations less than $m_{\text{eff,}\rho}^{-1} \sim 0.1$mm.  Furthermore, the chameleon ``sees'' only a shell of thickness $m_{\text{eff,}\rho}^{-1}$, so the experiment must use test masses not much larger than $m_{\text{eff,}\rho}^{-1}$.  Finally, the experiment must be sensitive enough to detect fifth forces with $\alpha \sim 1$.  Given these constraints, the E\"ot-Wash experiment, at the University of Washington, is a promising tool for detecting the chameleon.


\section{Chameleon predictions for the E\"ot-Wash Experiment} \label{predicts}

\subsection{The E\"ot-Wash experiment}
\begin{figure}[tb]
\includegraphics[width=2.0in]{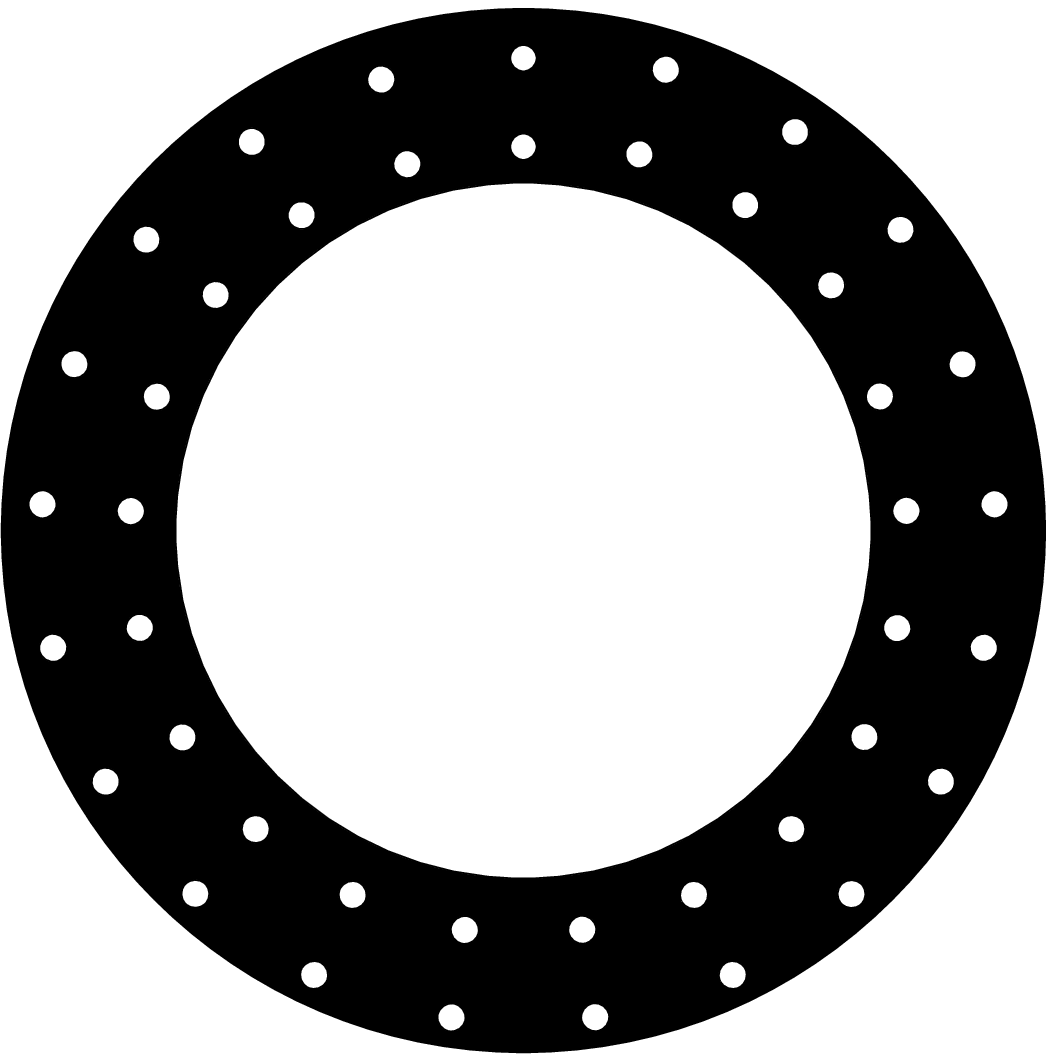}
\includegraphics[width=2.0in]{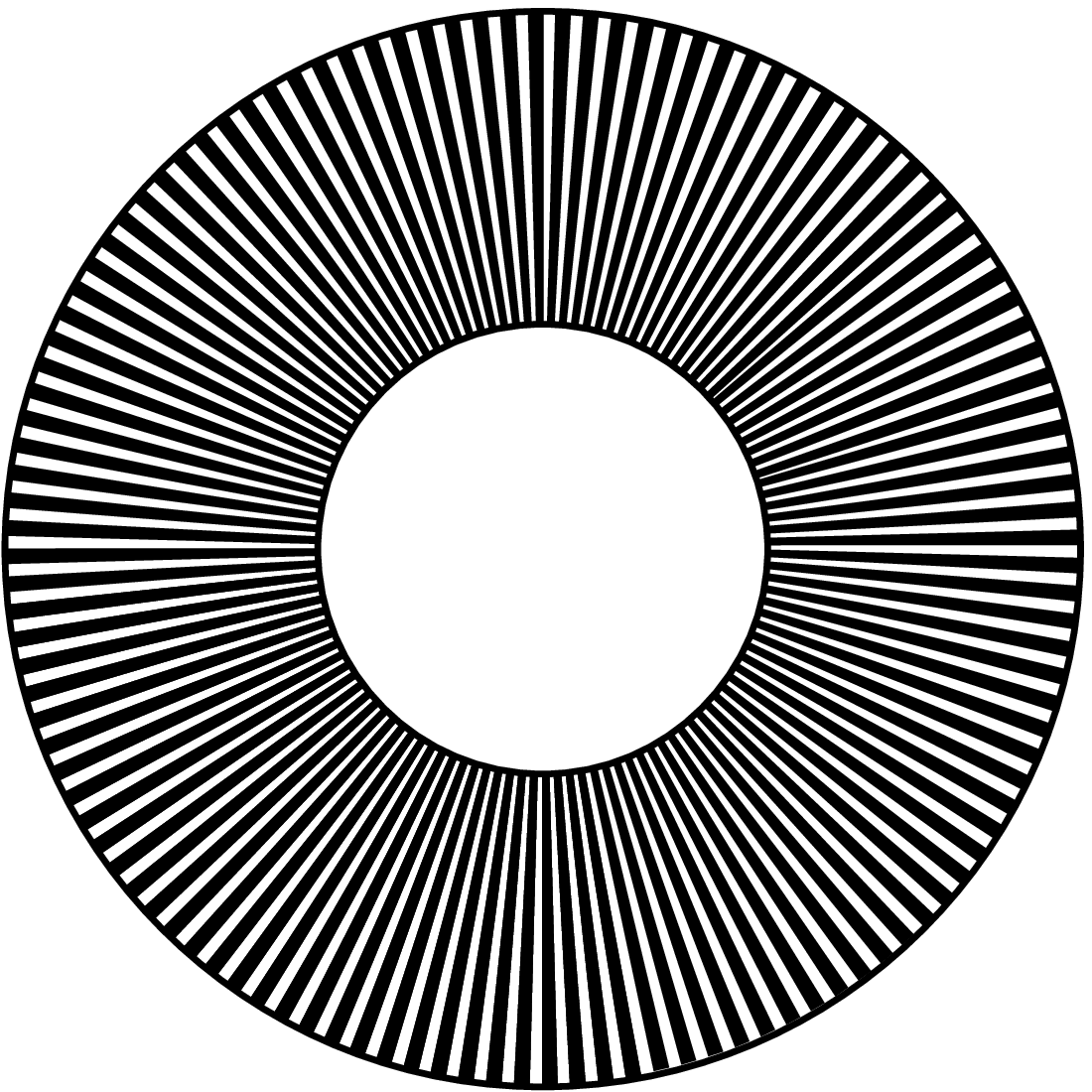}
\caption{Sketches of the current (left) and next-generation (right) E\"ot-Wash disks (not to scale).  \label{fig:EotWashApparatus}}
\end{figure}

The E\"ot-Wash experiment \cite{hoyle} uses two parallel disks to search for short-range deviations from the gravitational inverse square law.  The upper disk serves as a torsion pendulum, and the lower disk, the ``attractor'', rotates slowly below the pendulum.  In the current E\"ot-Wash experiment, each of the disks has 42 holes in it, at regular intervals, as sketched in Fig.~\ref{fig:EotWashApparatus}(left).  As the holes in the attractor disk rotate past those in the pendulum, the pendulum experiences a torque that tends to line up the two sets of holes.  By comparing the torque on the pendulum to that expected for purely Newtonian gravity, E\"ot-Wash can search for deviations from the inverse square law.

Although work is in progress using the current apparatus, the E\"ot-Wash group has already begun to construct a next-generation apparatus.  Rather than a series of holes on the pendulum and attractor disks, each of the next-generation disks will have a 120-fold symmetric pattern of wedges, as shown in Fig.~\ref{fig:EotWashApparatus}(right).  That is, each disk will resemble a pie carved into 240 equal slices, with every other slice removed, and with a circular region excised from the center.

\subsection{Solving the field equations}
The equation of motion of $\phi$ cannot be solved exactly for complicated density configurations such as the E\"ot-Wash pendulum and attractor disks.  In order to predict the form of the chameleon fifth force observable by E\"ot-Wash, numerical computations must be used.
The approach used here is to discretize space into a three-dimensional lattice, $\{(x,y,z)\}\rightarrow \{(x_i,y_j,z_k)| 1 \leq i \leq N_x, 1 \leq j \leq N_y, 1 \leq k \leq N_z\}$.  The field $\phi(x,y,z)$ is replaced by the quantities $\phi_{ijk}=\phi(x_i,y_j,z_k)$.  On this lattice, the approximate Hamiltonian is a function of the $\phi_{ijk}$s,
\begin{equation}
H \approx \sum_{i,j,k} {\Bigg [}
	\frac{1}{2}\left( \left(\frac{\Delta\phi}{\Delta x}\right)^2
			+ \left(\frac{\Delta\phi}{\Delta y}\right)^2
			+ \left(\frac{\Delta\phi}{\Delta z}\right)^2
		\right)
	+ \frac{1}{2} m_\phi^2 \phi^2_{ijk} + \frac{\lambda}{4!}\phi^4_{ijk}
	- \beta \rho_{ijk}\phi_{ijk} {\Bigg ]} 
	\Delta x \Delta y \Delta z\,,
\label{eqn:ham_lat}
\end{equation}
where $\rho_{ijk} = \rho(x_i,y_j,z_k)$, and $\Delta\phi/\Delta x$, etc. are numerical derivatives.  

Recall that, for static mass distributions, the field which minimizes the Hamiltonian also solves the field equation (\ref{eqn:eom}).  If the lattice has $N_x$, $N_y$, and $N_z$ points in the $x$, $y$, and $z$ directions, respectively, then the Hamiltonian is a function of $N_x N_y N_z$ variables.  The gradient of $H$ can be calculated by differentiating with respect to reach $\phi_{ijk}$, so $H$ may be minimized using a conjugate gradient algorithm.  Initial conditions are specified by choosing each $\phi_{ijk}$ from a uniform random distribution between $0$ and $\phi_\rho$ for the material making up the pendulum and attractor disks.

\begin{figure}[tb]
\includegraphics[width=3in]{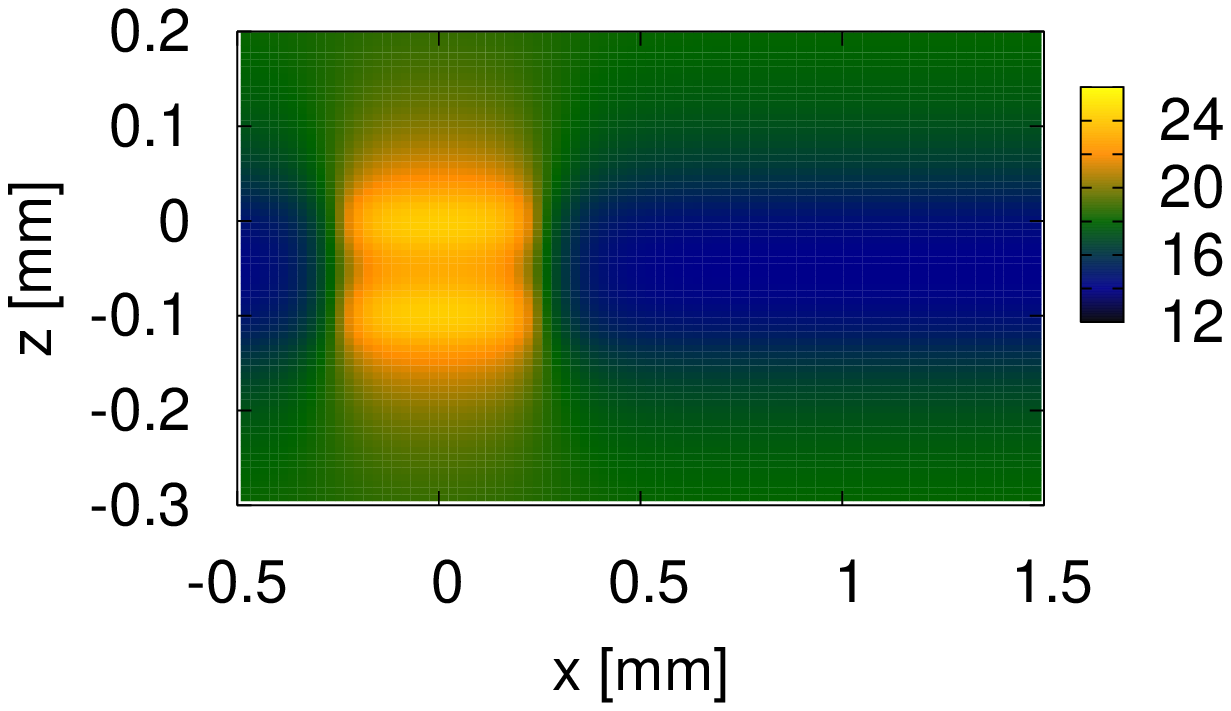}
\includegraphics[width=3in]{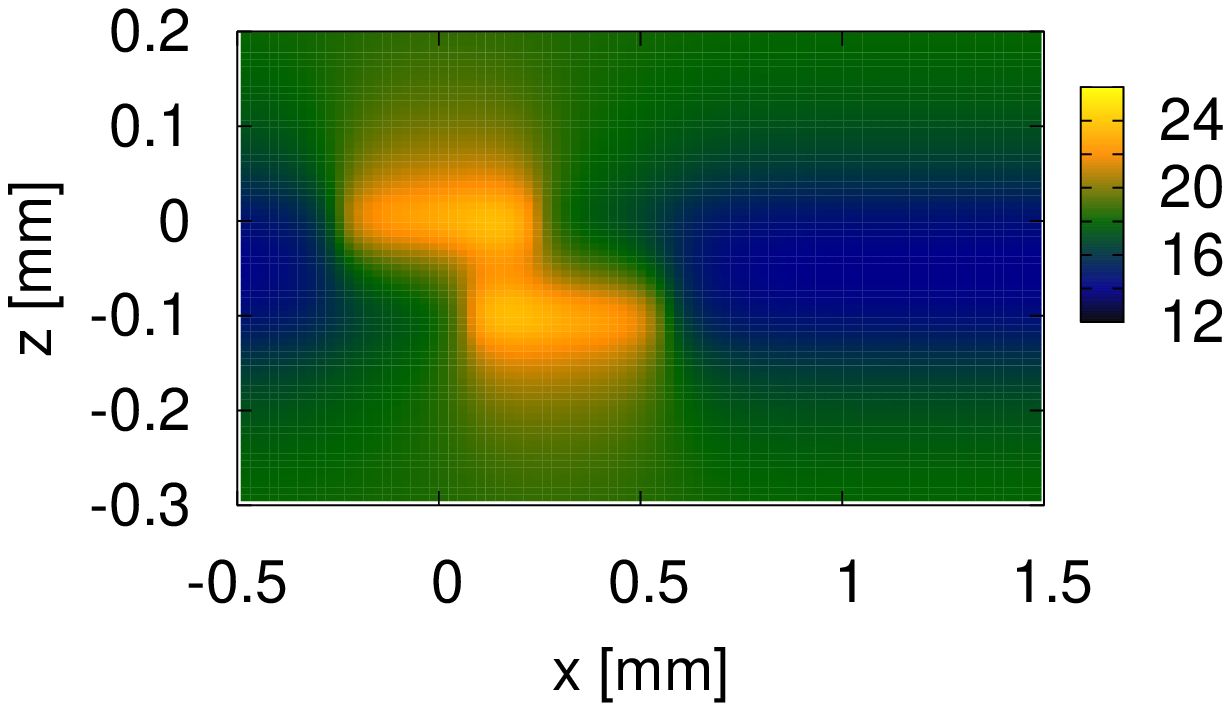}
\includegraphics[width=3in]{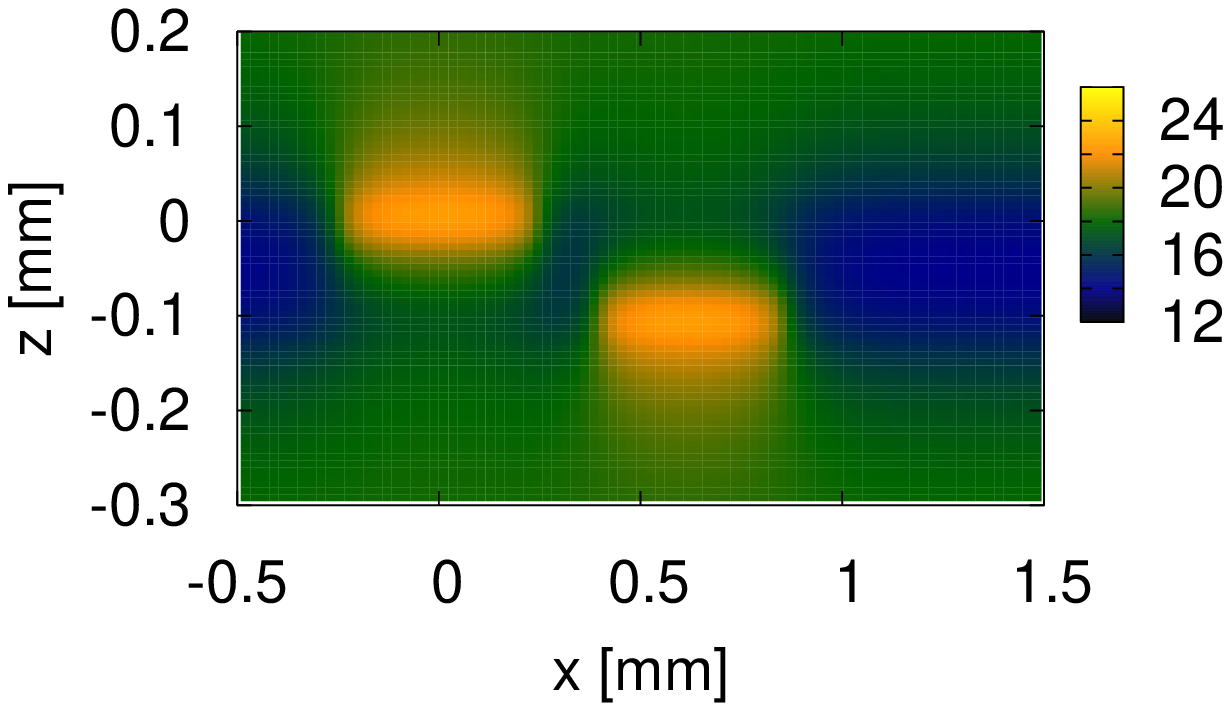}
\includegraphics[width=3in]{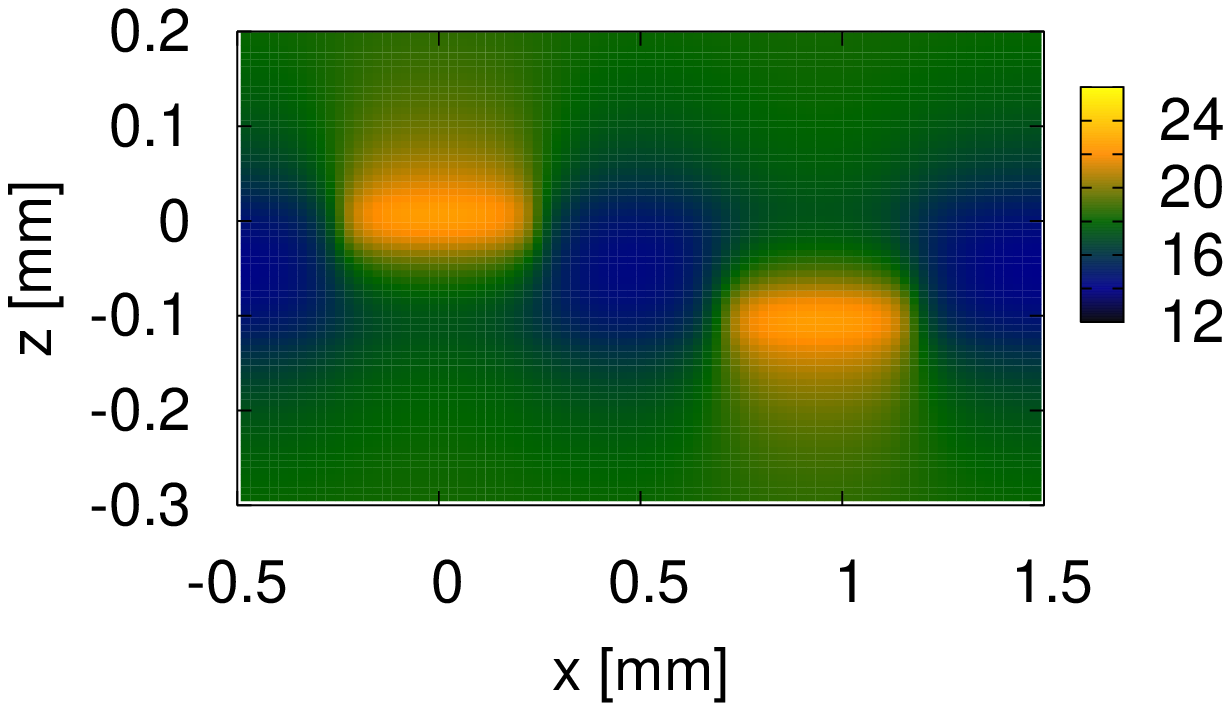}
\caption{Contour plots of an attractor wedge moving past a pendulum wedge.  The coordinates $x$ and $z$ represent the tangential and vertical directions, respectively, and the coloring represents the value of $\phi(x,z)$ in mm$^{-1}$.  The four plots correspond to rotation angles of $\theta=0$ (top left), $\theta=\pi/180$ (top right), $\theta=\pi/90$ (bottom left), and $\theta=\pi/60$ (bottom right).  \label{fig:movie}}
\end{figure}

In order to simplify the computation, we consider one single hole (or wedge) on the attractor disk moving past one hole (or wedge) on the pendulum disk at a time.  Also, in the case of the next-generation apparatus with 120 wedges, we replace the wedges by rectangular slabs, giving the problem more symmetry and speeding up the computations.  Contour plots of the field for the next-generation apparatus, with the attractor at several different angles with respect to the pendulum, are shown in Fig.~\ref{fig:movie}.

From these computations, the force $F(\theta)$ of one attractor hole (or wedge) on one pendulum hole (or wedge) can be computed as a function of angle.  The next attractor hole is at an angle $\theta_0$ relative to the first hole, and exerts a force in the opposite direction.  Neglecting any nonlinear effects between these two holes, which are separated by many times the length scale $m_{\text{eff,}\rho}^{-1}$, the force on the pendulum due to both holes is $F(\theta)-F(\theta_0-\theta)$.  Multiplying by the radius, and by the number of pendulum holes, gives the torque on the pendulum as a function of angle.

\subsection{Approximations in the computation}

\linespread{1.0}
\begin{table}[tb]
\begin{tabular}{|l|c|}
\hline
approximation 				& fractional error introduced\\
\hline
thin foil layer				& 0.039\\
decrease lattice spacing		& 0.013\\
force from nearby wedges		& 0.0042\\
wedges vs. rectangular slabs		& 0.0011\\
two attractor wedges (nonlinear)	& 0.00021\\
spurious torque at $\theta=0$		& 0.00002\\
change random number seed		& 0.00001\\
\hline
total					& 0.041\\
\hline
\end{tabular}
\caption{Summary of errors introduced due to approximations used in the computation. \label{tab:errors}}
\end{table}
\linespread{1.5} 

\begin{figure}[tb]
\includegraphics[width=3in]{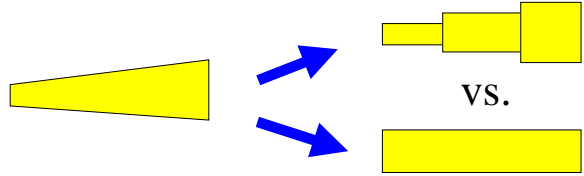}
\caption{Two different approximations to the wedge.\label{fig:err3}}
\end{figure}

Table \ref{tab:errors} summarizes the errors introduced into the final computed torque, in the next-generation E\"ot-Wash apparatus, by various computational approximations.  The two largest are the following:
\begin{itemize}
\item {\em thin foil layer.} The largest error comes from neglecting the thin layer of BeCu foil between the pendulum and attractor disks, which is necessary to isolate the two disks electrically.  Including this layer in the computation lowers the torque by about four percent.  
\item {\em decrease lattice spacing.}  Halving the lattice spacing in each direction changes the computed torque by approximately one percent.  
\end{itemize}
All other approximations lead to errors of less than one percent:
\begin{itemize}
\item {\em force from nearby wedges.}  Forces on the pendulum wedge due to all but the two closest attractor wedges were neglected.  An estimate of the error introduced is the force due to the third closest attractor wedge, whose magnitude is just $0.4\%$ of the combined force due to the first two wedges.
\item {\em wedges vs. rectangular slabs.} Our computation approximated each wedge in the next-generation apparatus as a rectangular slab.  This approximation may be improved by using three slabs, with widths corresponding to the inner, central, and outer radii of the disk, as shown in Fig.~\ref{fig:err3}.  The error listed in Table~\ref{tab:errors} is the difference between these two approximations.
\item {\em two attractor wedges (nonlinear contribution).} Since the field equation is nonlinear, one may object to our approximation of the force from two wedges, $F_2(\theta) \approx F(\theta)-F(\theta_0-\theta)$.  However, since the two are separated by a distance much greater than $m_{\text{eff,}\rho}^{-1}$, this approximation should introduce only a tiny error; this expectation was verified computationally.
\item {\em spurious torques.} We tested for the convergence of the Hamiltonian minimization by computing the residual torque at $\theta=0$, and by measuring the change in torque when initial conditions were chosen using a new random number seed; both spurious torques were negligible.
\end{itemize}
Combining all of the above errors in quadrature results in a total error of four percent.

We note that the above error estimates apply to the region of parameter space around $\beta \sim 1$ and $\lambda \sim 1$.  Ref. \cite{mota} points out that, for $\beta$ a few orders of magnitude greater than unity, the length scale $m_{\text{eff,}\rho}^{-1}$ becomes smaller than the thickness of the foil layer between the pendulum and attractor.  This allows the field to reach its maximum value inside the foil, so that the foil screens any variations in the fifth force on the pendulum as the attractor disk rotates.  Therefore, the E\"ot-Wash experiment will be insensitive to chameleon scalars with $\beta$ a few orders of magnitude greater than unity.

\subsection{Computed torques}

E\"ot-Wash will Fourier transform the measured torque, $N(\theta) = \sum N_n \sin(n\theta)$, and will report the first three Fourier coefficients not required to be zero by symmetry, $N_{J}$, $N_{2J}$, and $N_{3J}$.  Here, $J$ is the degree of rotational symmetry of the pendulum and attractor disks; $J=21$ for the current apparatus, and $J=120$ for the next-generation apparatus.  $N_{J} \sin(J\theta)$ is simply a sine wave with a period equal to $\theta_0=2\pi/J$, the angular distance between adjacent wedges.  Adding $N_{2J} \sin(2J\theta)$, with $N_{2J}\ll N_{J}$, shifts the first peak of the sine curve to the left, and adding $N_{3J} \sin(3J\theta)$, with $N_{3J} \ll N_{120}$, flattens the top of the sine curve.  Thus E\"ot-Wash will be sensitive to two features of the shape of the torque curve $N(\theta)$, in addition to its amplitude.

\begin{figure}[tb]
\includegraphics[width=3in]{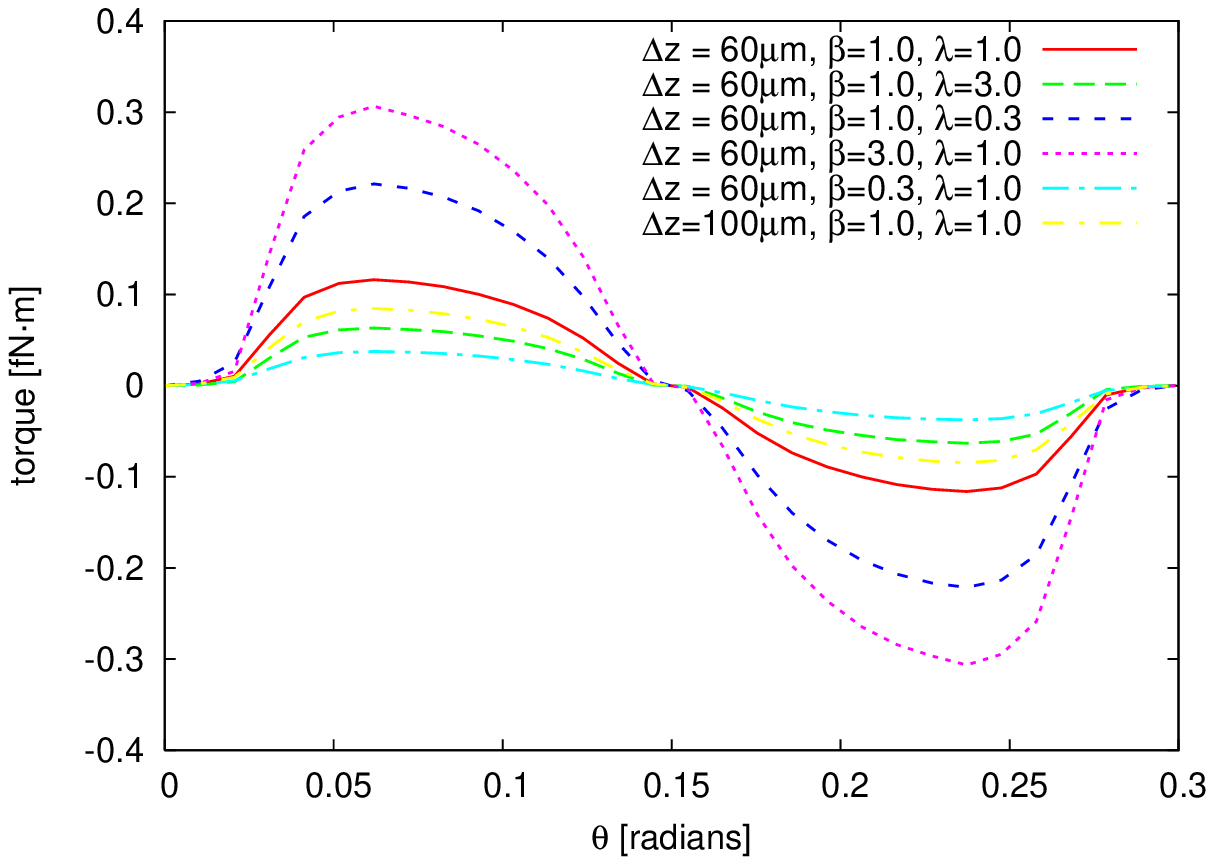}
\includegraphics[width=3in]{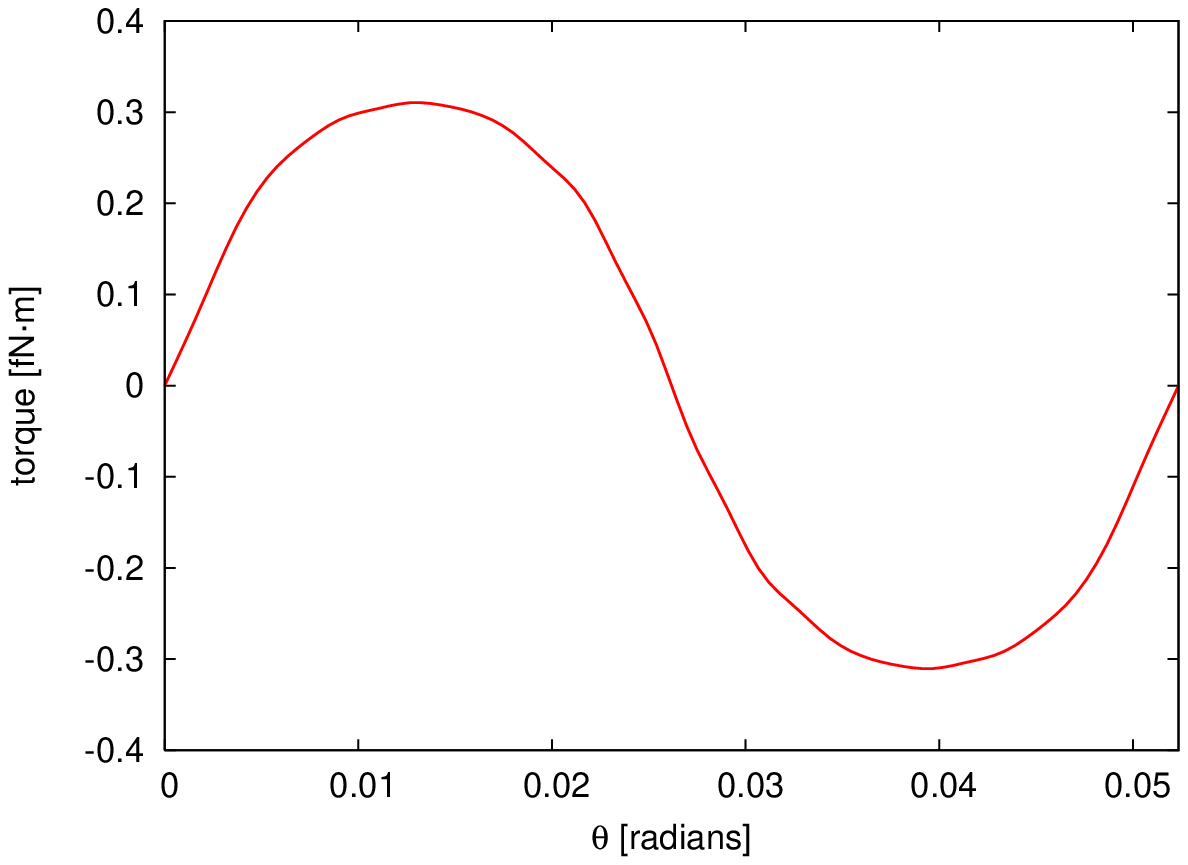}
\caption{(left) Torque vs. rotation angle for current (42 hole) E\"ot-Wash apparatus.  The torque is shown for several values of $\beta$, $\lambda$, and pendulum-attractor separation $\Delta z$.  (right) Torque vs. rotation angle for next-generation (120 wedge) E\"ot-Wash apparatus, with $\beta=\lambda=1$ and $\Delta z = 60$~$\mu$m. \label{fig:N_021_120}}
\end{figure}

The torque for the $42$-hole apparatus is shown in Fig.~\ref{fig:N_021_120}(left), for various values of the pendulum-attractor separation $\Delta z$, and the couplings $\beta$ and $\lambda$.   E\"ot-Wash is expected to probe separations as low as $\Delta z = 60\mu $m.  Since the E\"ot-Wash uncertainty in torque at these values of $\Delta z$ is approximately $0.1$ fNm, the chameleon with $\beta=\lambda=1$ will be invisible to this apparatus.  However, the primary Fourier coefficient scales as $N_{21} = 0.11 \beta^{0.91} \lambda^{-0.55}$ near $\beta=\lambda=1$.  If $\beta$ is larger than unity by a factor of a few, or if $\lambda \approx 1/10$, then E\"ot-Wash should be able to detect the chameleon.  

\begin{figure}[tb]
\includegraphics[width=3in]{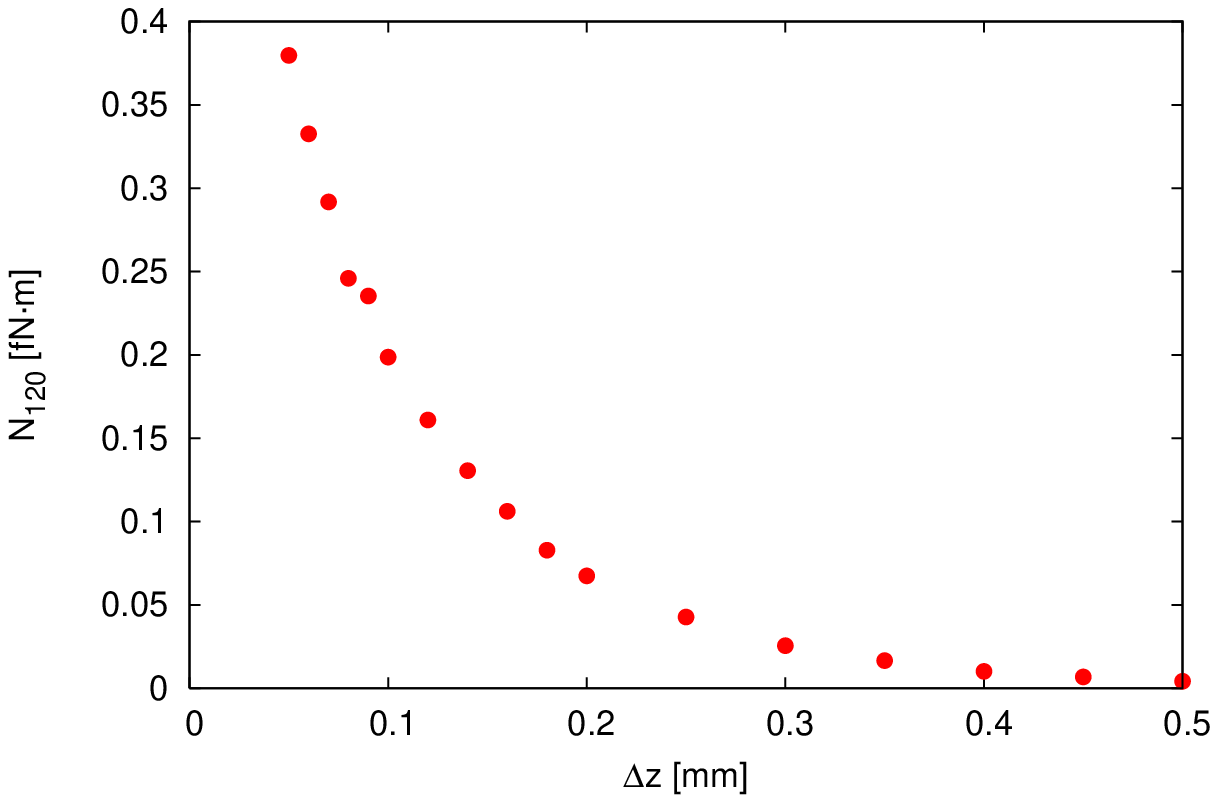}
\includegraphics[width=3in]{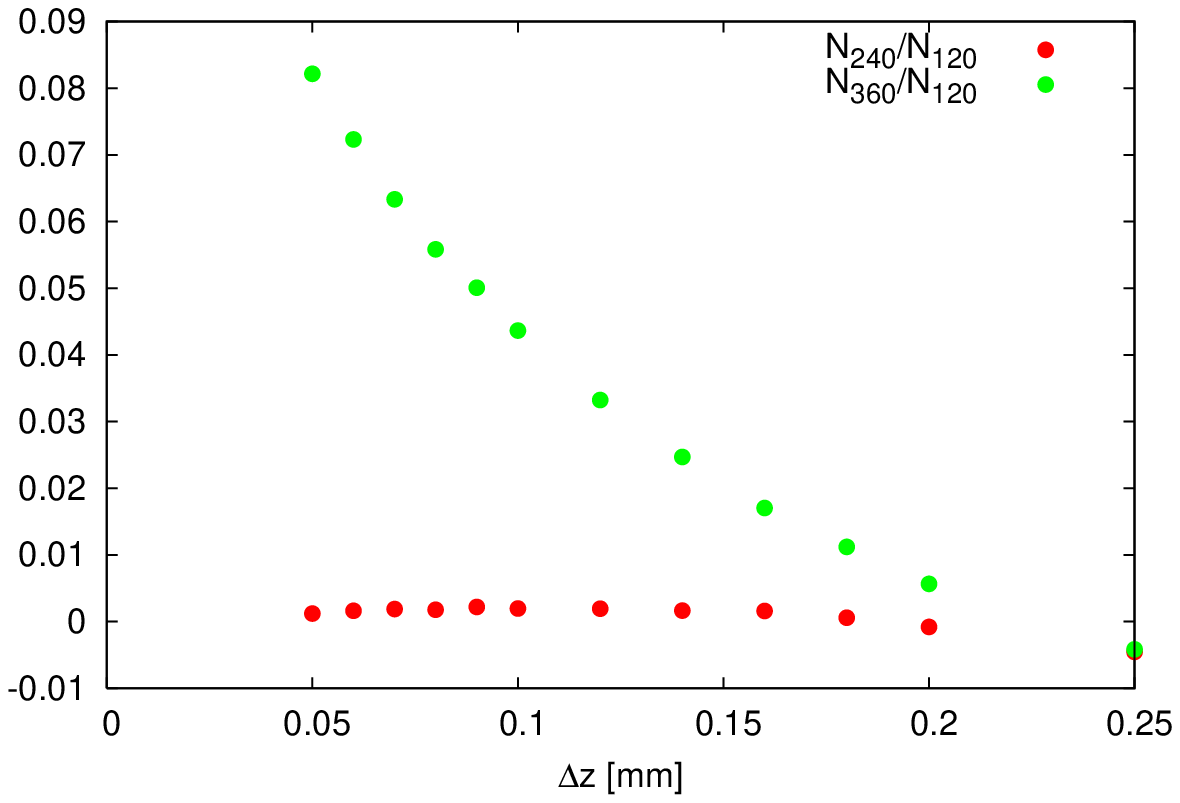}
\caption{Fourier coefficients of the torque vs. $\Delta z$ for the next-generation apparatus.  (left) $N_{120}$.  (right) $N_{240}/N_{120}$ and $N_{360}/N_{120}$. \label{fig:fourier}}
\end{figure}

The next-generation E\"ot-Wash apparatus, with 120 wedges, will be several times more sensitive to the chameleon-mediated fifth force.  This should allow the detection of a chameleon with $\beta=\lambda=1$ at the $3\sigma$ level, as shown in Fig.~\ref{fig:N_021_120}(right).  It is evident from the plot of $N(\theta)$ in Fig.~\ref{fig:N_021_120}(right) that the torque is approximately a sine wave $\sin(120\theta)$ with a flattened top.  The three Fourier coefficients are shown in Fig.~\ref{fig:fourier} as functions of $\Delta z$.  Note, first, that $N_{120}$ falls off rapidly with increasing $\Delta z$.  The chameleon will be detectable at the $2\sigma$ level only below $\Delta z \approx 0.1$mm.  Also note that the ratio of $N_{360}$ to $N_{120}$ drops with increasing $\Delta z$.  That is, the peak becomes less flattened with increasing separation.  Unfortunately, $N_{360}$ is too small to be detectable for $\beta=\lambda=1$.  The two Fourier coefficients scale with the coupling constants as $N_{120} = 0.38 \beta^{1.34}\lambda^{-0.33}$~fNm and $N_{360} = 0.031 \beta^{1.68} \lambda^{-0.16}$~fNm at $\Delta z=0.05$~mm, and $N_{120} = 0.20 \beta^{1.22}\lambda^{-0.39}$~fNm and $N_{360} = 0.0087 \beta^{1.68} \lambda^{-0.16}$~fNm at $\Delta z=0.1$~mm, so there is some possibility that this peak flattening will be seen for large $\beta$.


\section{Conclusion} \label{conclu}

We have shown that a chameleon field with a quartic self interaction mediates an attractive force, which falls off rapidly with separation between two massive objects, and is sensitive only to the outer shell of matter in a large object.  Given the size of its test masses and the length scales that it probes, the E\"ot-Wash experiment is a promising instrument for searching for this fifth force.  
For unit values of the matter coupling constant $\beta$ and the quartic coupling constant $\lambda$, the current E\"ot-Wash apparatus is not expected to find evidence for the chameleon fifth force.  However, the experiment is capable of constraining a very interesting region of parameter space; the expected chameleon signal $N_{21} = 0.11 \beta^{0.91} \lambda^{-0.55}$ will be detectable for certain values of $\beta$ and $\lambda$ not much different from unity.

The next-generation E\"ot-Wash experiment will be several times more sensitive to chameleon-mediated fifth forces.  We expect E\"ot-Wash to detect, or to rule out, a chameleon force with unit couplings at the $3\sigma$ level.  If a small-scale deviation from Newtonian gravity is observed, this deviation may be compared to our predictions for $\beta=\lambda=1$ in Fig.~\ref{fig:fourier}.  The torque Fourier coefficients at other values of the coupling constants, and at a pendulum-attractor separation of $\Delta z = 0.05$mm, are $N_{120} = 0.38 \beta^{1.34}\lambda^{-0.33}$ fNm and $N_{360} = 0.031 \beta^{1.68} \lambda^{-0.16}$ fNm.  Thus, if $\beta$ is large enough for the chameleon force to be detected by the current E\"ot-Wash experiment, then the next-generation experiment should detect a peak flattening $N_{360} > 0$.

If a short-range fifth force is observed by the E\"ot-Wash experiment, the next goal will be to distinguish between chameleon and Yukawa fifth forces.  Although unit-strength Yukawa fifth forces have already been ruled out by E\"ot-Wash, a Yukawa force tuned to be a few orders of magnitude smaller than gravity could conceivably fit the observed fifth force.  As argued in Sec. \ref{subsec:chameleonmech}, chameleon and Yukawa forces that are identical at small scales begin to differ substantially at length scales of the order of $m_{\text{eff,}\rho}^{-1}$.  Unfortunately, a chameleon force with $\beta=\lambda=1$ is only visible over a range of pendulum-attractor separations $0.05\textrm{mm} \leq \Delta z \leq 0.1\textrm{mm}$, a range whose width is approximately equal to the length scale $m_{\text{eff,}\rho}^{-1} = 0.047\textrm{mm}$ inside the pendulum and attractor wedges.  Distinguishing between a chameleon force and a Yukawa force will be challenging, and unless $\beta$ is sufficiently greater than unity, a more sensitive experiment will be necessary.

\section*{Acknowledgments}
We are indebted to E.~Adelberger, T.~Cook, S.~Hoedl and D.~Kapner of the E\"ot-Wash group for frequent help with the details of the current and future apparatus. We are grateful to P.~Steinhardt for collaboration in the early stages of this work and many insightful discussions. We thank P.~Brax, C.~Hirata, J.~Moffat, D.~Mota and A.~Nelson for helpful comments.  
The work of A.~U.\ was supported in part by the Department of Energy under Grant No.\ DE-FG02-91ER40671, by a National Science Foundation Graduate Research Fellowship, and by the Perimeter Institute for Theoretical Physics.
The work of S.~G.\ was
supported in part by the Department of Energy under Grant No.\ DE-FG02-91ER40671, and by the Sloan Foundation.
The research of J.~K.\ at Perimeter Institute is supported in part by the
Government of Canada through NSERC and by the Province of Ontario through
MEDT.


\end{document}